\documentclass[12pt, citesort]{iopart}
\usepackage[numbers,sort&compress]{natbib}

\usepackage{graphicx}
\usepackage{iopams}
\eqnobysec
\usepackage{color}

\begin{document}
\title {Electronic and optical properties of ternary kagome Rb$_{2}$Ni$_{3}$S$_4$: A density functional study}

 \author{Gang Bahadur Acharya$^{1,2}$, Se-Hun Kim$^{3}$, and Madhav Prasad Ghimire$^{1,2,3}$} 

 \address{$^1$~Central Department of Physics, Tribhuvan University, Kirtipur, 44613 Kathmandu, Nepal}
%\address{$^2$~Condensed Matter Physics Research Center, Butwal-11, 32907 Butwal, Nepal}
 \address{$^2$~Leibniz Institute for Solid State and Materials Research - IFW Dresden, Helmholtzstr. 20, 01069 Dresden, Germany}
 \address{$^3$~Faculty of Science Education, Jeju National University, Jeju 63243, Republic of Korea}

\ead{madhav.ghimire@cdp.tu.edu.np}
\vspace{10pt}
%\begin{indented}
%\item[]September 2023
%\end{indented}
 
\begin{abstract}
 The application of semiconductors with optical properties has grown significantly in the development of semiconductor photovoltaics. Here, we explore the electronic and optical properties of ternary transition metal sulfide Rb$_{2}$Ni$_{3}$S$_4$ by means of density functional theory. From the structural perspective, Ni atoms are found to form a kagome-like lattice in a two-dimensional plane of Rb$_{2}$Ni$_{3}$S$_4$. 
 From our calculations, Rb$_{2}$Ni$_{3}$S$_4$ is found to be a semiconductor with an indirect band gap of $\sim$0.67 eV. Strong hybridization was observed between the S-3\textit{p} with the Ni-3$d_{xz}$ and Ni-3$d_{yz}$ orbitals. Interestingly, a flat band was noticed below the Fermi level demonstrating one significant feature of kagome lattice. From the optical calculations, Rb$_{2}$Ni$_{3}$S$_4$ is found to exhibit optical activity in both the visible and lower ultraviolet regions of the incident photon energies. The optical response suggests this material may be a potential candidate for opto-electronic device, given its ability to interact with light across a broad range of wavelengths. This work is expected to motivate  the experimental group for transport measurements and may provide a new foundation in optics.

\end{abstract}
\maketitle
\ioptwocol

\section{Introduction}

 The family of ternary transition metal chalcogenides represented by the molecular formula \textit{A}$_{2}$\textit{M}$_{3}$\textit{X}$_4$ (where A includes K, Rb, Cs; M includes Ni, Pd, Pt; and X includes S, Se) demonstrates diverse crystal symmetries \cite{bronger1991schichtstrukturen}. These compounds possess unique properties, featuring a quasi-two dimensional (2D) layered structure, where the transition metal elements create kagome nets together with chalcogenide atoms. 
 Kagome compounds are widely recognized as excellent platforms for exploring novel topological properties. The specific arrangement of atoms in a kagome pattern leads to remarkable transport characteristics, including negative magneto-resistance, anomalous Hall conductivity, and Fermi arc surface states \cite{kang2020dirac,chen2021large,liu2018giant,
ghimire2019creating,yang2017topological,kubler2018weyl,liu2017anomalous,ren2022plethora}. 
They are also found to show interesting optical features which can give rise to optical device fabrication.
The recently discovered kagome CsV$_3$Sb$_5$ was reported to exhibit efficient absorption of ultraviolet radiation and has a strong reflectivity for visible light \cite{naher2023comprehensive}. 
Likewise, KV$_3$Sb$_5$ has been reported as a potential candidate for plasmon-mediated hot carrier applications in the infrared region \cite{he2022kagome}.  Kagome Nb$_3$Cl$_8$ was reported to be a semiconductor which shows large optical anisotropy ~\cite{bouhmouche2024highly}.

In the family of ternary chalcogenides Rb$_{2}$Pd$_{3}$Se$_4$, Li \textit{et al}.  predicted the  superconductivity by applying pressure wherein Lifshitz transition was observed \cite{li2022superconductivity}.  Elder \textit{et al}. performed the magnetic and transport measurements in K$_2$Ni$_3$S$_4$ suggesting the possibility of co-existence of Ni$^{1+}$ and Ni$^{2+}$ in the crystal \cite{elder1996structural}.
Likewise, we recently studied the electronic and optical properties of Cs$_{2}$Ni$_{3}$S$_4$ in addition to the effect of vacancy defects in the crystal. Our study suggests this material to be a suitable candidate for optoelectronic device \cite{acharya2023electronic}. 
While studies of Rb$_2$Ni$_3$S$_4$ are uncommon, the existing results suggest that the material could be interesting. 
For instance, Bronger \textit{et al}. first synthesized and  carried out the magnetic susceptibility measurements above 90 K to explore magnetism in Rb$_{2}$Ni$_{3}$S$_4$. They did not find any evidence of magnetism \cite{bronger1970thio}. 
At 4.2 K, the magnetism of Rb$_2$Ni$_3$S$_4$ indicates that the ordered phase may represent the weak ferromagnetic phase \cite{kato1988}.
Fukamachi \textit{et al}. reported the observation of weak ferromagnetism and semi-conducting characteristics in Rb$_2$Ni$_3$S$_4$. However, their findings, based on both nuclear magnetic resonance and magnetic measurements, led them to propose that Rb$_2$Ni$_3$S$_4$ is a band insulator. Consequently, they suggested that the intrinsic property of this system to be non-magnetic \cite{fukamachi1999magnetization}. Later in 2002, Hondou  \textit{et al}. 
performed DFT calculations and find that the material is a semiconductor, featuring flat bands below the Fermi level (\textit{E}$_{\mbox{F}}$) \cite{hondou2002reinvestigation}. Additional investigations using photoemission spectroscopy have unveiled the moderate electron correlation and substantial hybridization between Ni-3\textit{d} and S-3\textit{p} orbitals, in agreement with the results from the band structure calculations \cite{nawai2004electronic}. 
Experiments have been performed in order to examine the origin of magnetism in water-immersed Rb$_2$Ni$_3$S$_4$. A possible ferromagnetic phase induced by the partial collapse of crystal structure was obtained. The intrinsic substance is non-magnetic semiconductor. The presence of a planar sulfur coordination around Ni$^{2+}$ results in a low spin state with Ni in its 3\textit{d}$^8$ configuration  \cite{hondou2007attempt}. Despite a number of studies, there are no reports on optical properties of Rb$_2$Ni$_3$S$_4$. With an aim to fill this gap and to show the interesting behaviour of this material, we explore the detailed electronic structure and optical properties of Rb$_{2}$Ni$_{3}$S$_4$.

In this work, we report the electronic as well as optical properties of Rb$_{2}$Ni$_{3}$S$_4$. This material is found to be a non-magnetic semiconductor with an indirect band gap  of $\sim$0.67 eV. The optical activity in this material is prominent within the visible and lower ultraviolet spectrum.

\section{Structural and computational details}
\begin{figure*}[!htb]
 		\centering
 		\includegraphics[scale=0.25]{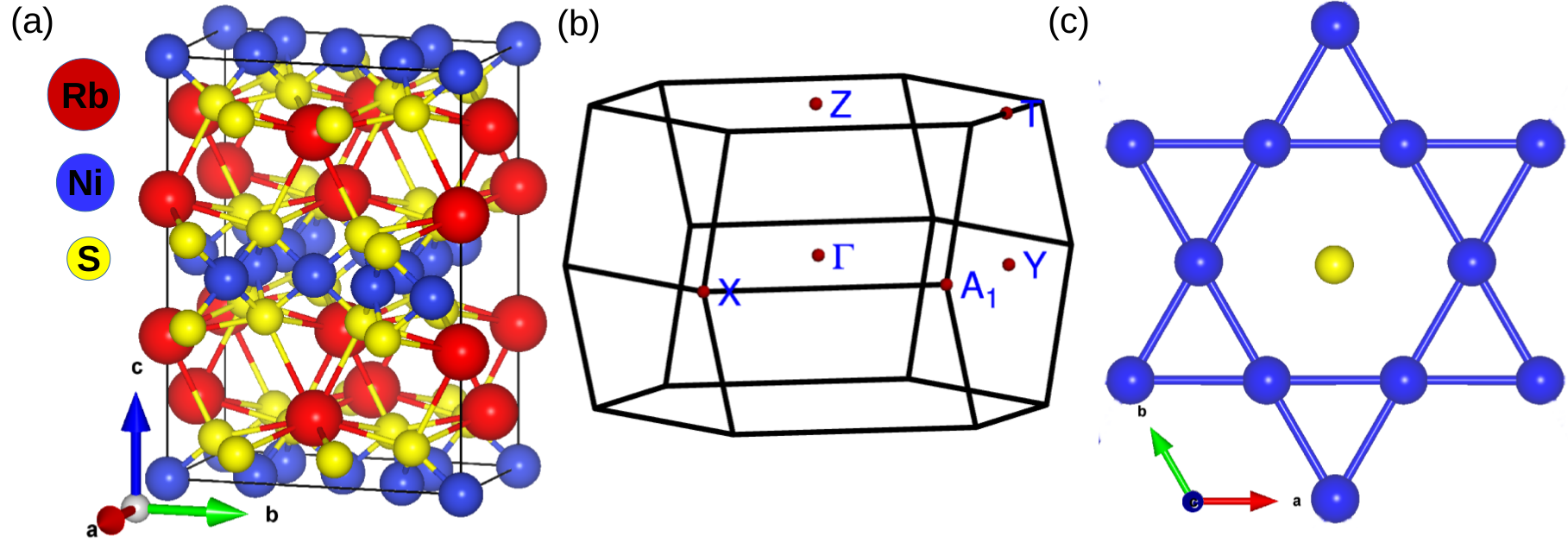} 		
 		\vspace{-0.2cm}
 		\caption{(a) Crystal structure of Rb$_{2}$Ni$_{3}$S$_4$. (b) Three dimensional Brillouin zone. Red dots are indicates the high symmetry points in momentum space. (c) Ni-ions constitute a kagome lattice in \textit{ab} plane.} 
 		\label{optical}
 	\end{figure*}

Rb$_{2}$Ni$_{3}$S$_4$ crystallizes in an orthorhombic crystal structure with space group \textit{Fmmm} (no. 69) and belongs to the point group \textit{D$_{2h}$}. The experimental lattice parameters used in this study are $a$ = 5.862 \AA,  $b$= 9.937 \AA \hspace{0.05cm} and $c$ = 13.758 \AA \hspace{0.05cm} \cite{bronger1991schichtstrukturen}.
The crystal system consists of four in-equivalent atoms labeled as Rb, Ni(I), Ni(II), and S, respectively. In Rb$_{2}$Ni$_{3}$S$_4$, Rb$^{1+}$ ions are surrounded by eight equivalent S$^{2-}$ ions in an 8-coordinate geometry. The Rb–S bond distances exhibit a range from 3.38 {\AA} to  3.45 \AA. All Ni$^{2+}$ ions are coordinated by four equivalent S$^{2-}$ ions, adopting a square co-planar arrangement (see Fig. 1 (a)). At the first and second Ni$^{2+}$ site, the Ni–S bond length measured are 2.22 \AA. 
%\textbf{underline comment: This all says "Rb2Ni3S4 is centrosymmetric", right?} 
 Furthermore, Rb$_{2}$Ni$_{3}$S$_4$ exhibits symmorphic crystal symmetry, and is centrosymmetric. Table 1 presents the fully relaxed atomic coordinates for Rb$_{2}$Ni$_{3}$S$_4$.
 %\rtext{My answer: yes}
 
In order to investigate the electronic and optical characteristics of selected compound, we employed the DFT approach \cite{kohn1965self} using the full potential local orbital (FPLO) code \cite{koepernik1999full}, version 18.00. The standard generalized gradient approximation (GGA) within the parameterization of Perdew, Burke, and Ernzerhof (PBE-96) \cite{perdew1996generalized}  was taken into account for the exchange and correlation energy. 
Self-consistent calculations were performed in the scalar relativistic and four component full relativistic modes of FPLO. In the full relativistic mode that includes spin-orbit coupling in all orders, the direction of magnetization was fixed by a global setting of the spin quantization axis.
 However, since the inclusion of spin-orbit coupling did not significantly impact the electronic properties, we focus our discussion only on the outcomes from the scalar-relativistic calculations. 
 For the optical property calculations, we have used the FOPTICS module of FPLO which is based on random phase approximation. 
 In the electronic structure calculations, a \textit{k}-mesh subdivision of $12\times12\times12$ was used to cover the full Brillouin zone. As for the optical property calculations, a finer \textit{k}-mesh subdivision of $32\times32\times32$ was employed. To ensure the accuracy of the energy, the convergence criteria for self-consistency as set to  10$^{-8}$ Hartree. During the structural optimization, a force convergence of 10$^{-3}$ eV/{\AA} was applied.

 Our Calculations done with FPLO are consistent with results performed using WIEN2k ~\cite{blaha2001wien2k}. The accuracy of FPLO can be well compared with other electronic structure codes as reported by Lejaeghere $et$ $al.$~\cite{lejaeghere2016reproducibility}.

\begin{table}[htbp]
\caption{Total number of in-equivalent atoms and their Wyckoff positions in Rb$_{2}$Ni$_{3}$S$_4$.}
\vspace{0.3cm}
\centering

\begin{tabular}{ccccc}
\hline
\hline
Atom & Point location&x&y&z\\
\hline
\hline
 Rb&8i&0&0&0.343\\
 \hline
Ni(I)&8e&-1/4&-1/4&0 \\
\hline
 Ni(II)&4a&0&0&0 \\
 \hline
 S&16m&0& 0.168&0.106\\
\hline
\hline
\end{tabular}
\label{tab:mytable}
\end{table}

\section{{Results and discussion}}

 \subsection{Electronic properties}
 \begin{figure}[]
\centering
  \hspace{-0.3cm}\includegraphics[scale=2.35]{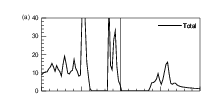} \\
         \vspace{-1cm}
     \hspace{-0.4cm}  \includegraphics[scale=2.35]{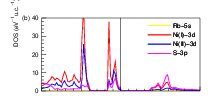} \\
       \vspace{-1.3cm}
  \includegraphics[scale=0.3]{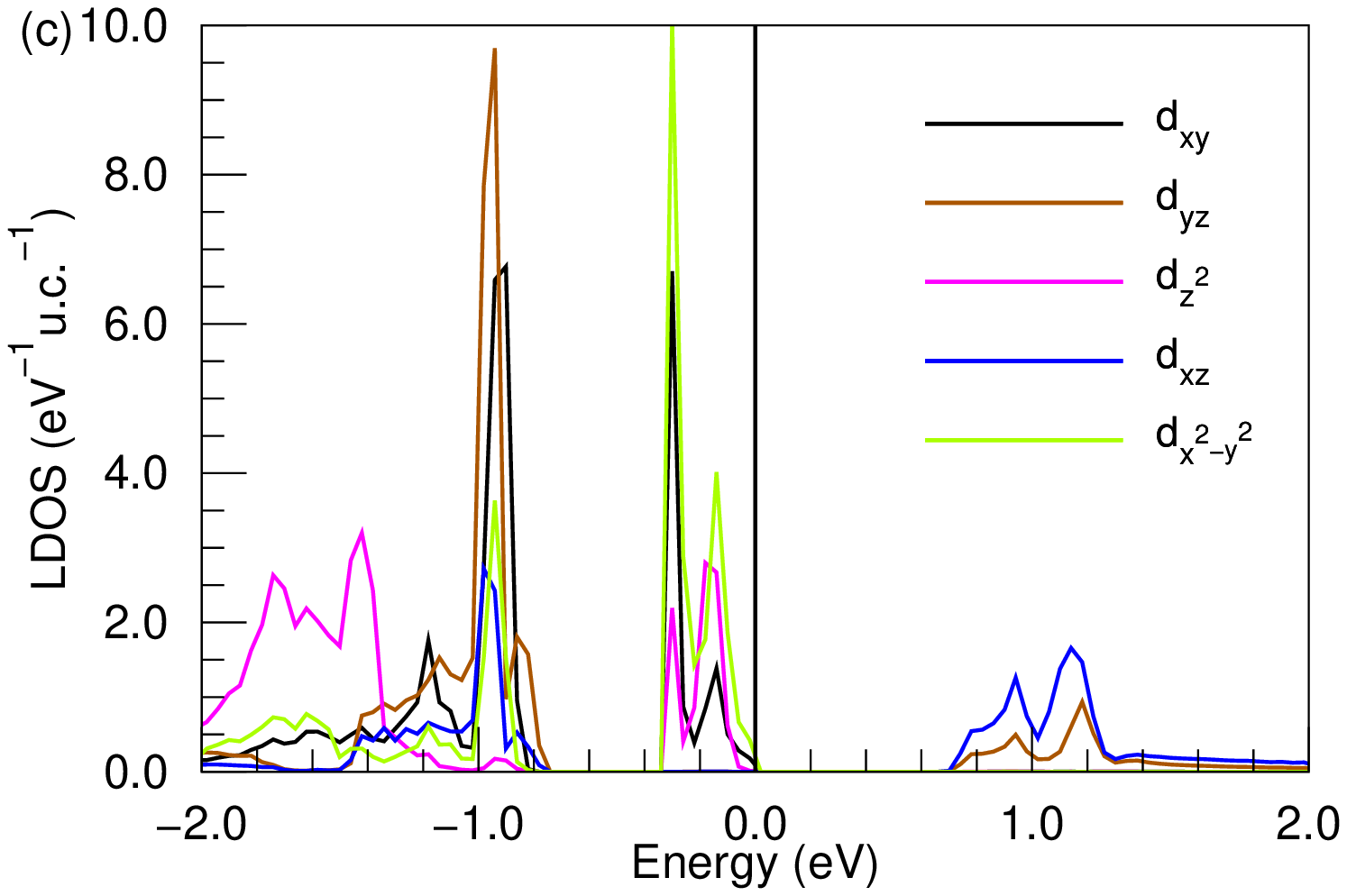} \\
 %  \vspace{-1.9cm}
 %  \includegraphics[scale=0.3]{RbNiS_Ni3d_site5_only_ldos.eps}\\
  % \vspace{-1.9cm}
  % \includegraphics[scale=0.3]{RbNiS_S3p_site6_only_ldos.eps}    
    \vspace{-0.5cm}
 \caption{Total, partial and local DOS of Rb$_2$Ni$_3$S$_4$. (a) Total DOS. (b) partial DOS. (c) Projected  local DOS for Ni(I)-3\textit{d}. The vertical solid line at zero energy denotes \textit{E}$_{\mbox{F}}$.}

   \label{optical}
 	\end{figure}
To  understand the electronic properties, we initiated our calculations by searching for the stable ground state of Rb$_2$Ni$_3$S$_4$. To achieve this, we performed both nonmagnetic and ferromagnetic calculations in the scalar relativistic case (without spin-orbit coupling). The outcome was a nonmagnetic ground state, attributed to the square planar coordination of Ni$^{2+}$ with S$^{2-}$ ions. Our DFT ground state result agrees with the experimental nonmagnetic ground state \citep{bronger1970thio,dehnicke2003zeitschrift}. 
Additionally, we performed full-relativistic calculations, considering spin-orbit coupling. However, this consideration did not significantly affect the electronic structure of Rb$_2$Ni$_3$S$_4$. Consequently, we focus solely on the scalar relativistic results throughout our discussions. 
 
Figure 2 (a, b, and c) illustrates the total, partial, and local density of states (DOS), respectively. In the total DOS, high peaks both above and below the \textit{E}$_{\mbox{F}}$ within a specific energy range result from the bonding and anti-bonding nature of the orbitals. Ni and S states are mainly dominate the DOS, while Rb has a minimal contribution (see Fig. 2 (b)). This arises due to the presence of only one electron in the outermost 5s orbital of Rb atom. 

To precisely understand the contributions of the orbitals, we analyze the partial and local DOS. 
At around 1.3 eV in the DOS plot (see Fig. 2),  antibonding states are observed which are contributed by Ni(I)-$d_{xz}$ and S-3\textit{p} states. This gives rise to peaks above \textit{E}$_{\mbox{F}}$.  While around -0.9 eV and -0.38 eV below \textit{E}$_{\mbox{F}}$ are contributed by Ni(I)-$d_{x^2-y^2}$ and $d_{xy}$ orbitals giving rise to peaks due to the bonding states.
The analysis of the local DOS projection (see Fig. 2 (c)) reveals that the insulating behavior partially originates from the crystal field splitting of the energy bands \cite{li2022superconductivity}. It is well-established that each nickel (Ni) ion is surrounded by four S ions, forming a square planar coordination arrangement (see Fig. 1 (a)). In this situation, the S$^{2-}$ ions strongly bind with the central Ni$^{2+}$ metal, causing a separation of energy levels known as crystal field splitting. As a result, metal ions with a d${^8}$ configuration tend to adopt a square planar geometry when exposed to a strong field. This arrangement forms low spin complexes where the eight d electrons fill the lower-energy $d_{xz}$, $d_{yz}$, $d_z^2$ , and $d_{xy}$ orbitals, while the high-energy $d_{x^2-y^2}$ orbital remains unoccupied. This also helps explain the origin of the nonmagnetic ground state even in the presence of a typically magnetic atom Ni. Above the \textit{E}$_{\mbox{F}}$, the width and magnitude of the DOS of S-3\textit{p} and Ni-3\textit{d} ($d_{xz}$ and $d_{yz}$ orbitals) (see in Fig. 2 (b)) are almost equal, suggesting strong hybridization within the energy range from 0.7 to 1.2 eV \cite{hondou2002reinvestigation}.
%a pronounced hybridization occurs between the S-3p orbitals with the Ni-3d Ni-3$d_{xy}$, $d_{x^2-y^2}$, and $d_z^2$ orbitals in the energy range spanning from 0 to -0.035 eV. Conversely, above \textit{E}$_{\mbox{F}}$, S-3\textit{p$_x$} and S-3\textit{p$_y$} orbitals exhibit significant hybridization with the Ni-3$d_{xz}$ and Ni-3$d_{yz}$ orbitals within the energy range from 0.7 to 1.2 eV (see Fig. 2(c-e)). 

Based on the analysis of the electronic band structure, Rb$_2$Ni$_3$S$_4$ should exhibit semiconducting behavior with an energy band gap of $\sim$0.67 eV. These findings are consistent with the previously reported DFT value of 0.66 eV \cite{PhysRevB.69.045103} and the experimental value of 0.80 eV deduced from a transport study \cite{hondou2002reinvestigation}. The nature of the band gap is indirect. This is attributed to the occurrence of the VBM at the high symmetry point X and the CBM at the T-Z line in the momentum space (see Fig. 3). Additionally, the kagome lattice of Ni$^{2+}$ leads to the formation of flat bands below \textit{E}$_{\mbox{F}}$, which are largely comprised of the Ni-3\textit{d} orbitals. 
Theoretical studies have suggested that such flat band systems in a kagome lattice can give rise to various exotic many-body phenomena when \textit{E}$_{\mbox{F}}$ is proximal to them \cite{ 
hondou2002reinvestigation,kang2020topological,wang2013competing,yu2012chiral}.
 \begin{figure}[!htb]
 		\centering
 \includegraphics[scale=0.3]{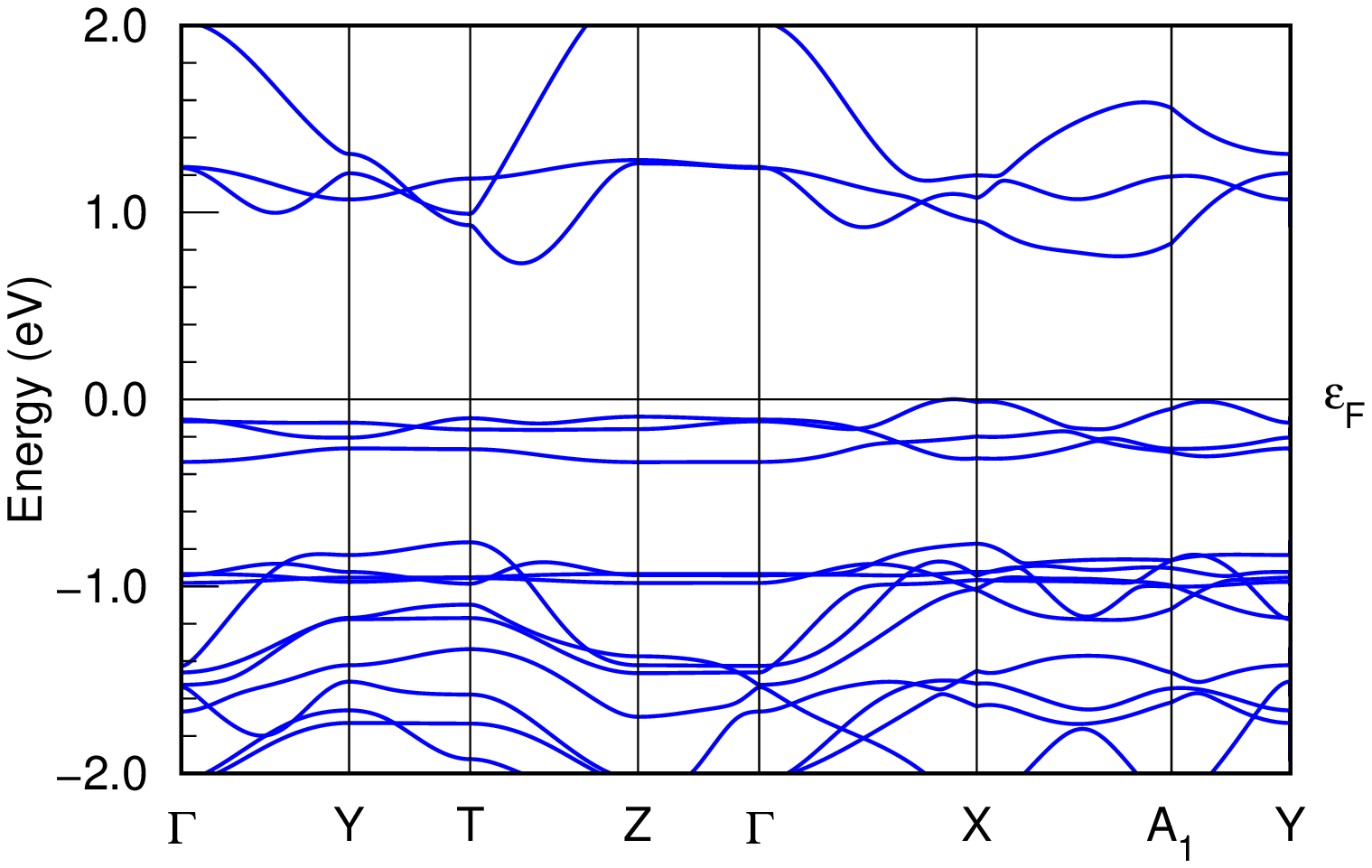} 
 		\vspace{-1.2cm}
 		\caption{Electronic band structure of Rb$_2$Ni$_3$S$_4$. The horizontal solid line at zero energy denotes \textit{E}$_{\mbox{F}}$.} 
 		\label{optical}
 	\end{figure}

 \subsection{Optical properties}
\begin{figure*}[!htb]
 		\centering
 	\includegraphics[scale=1.4]{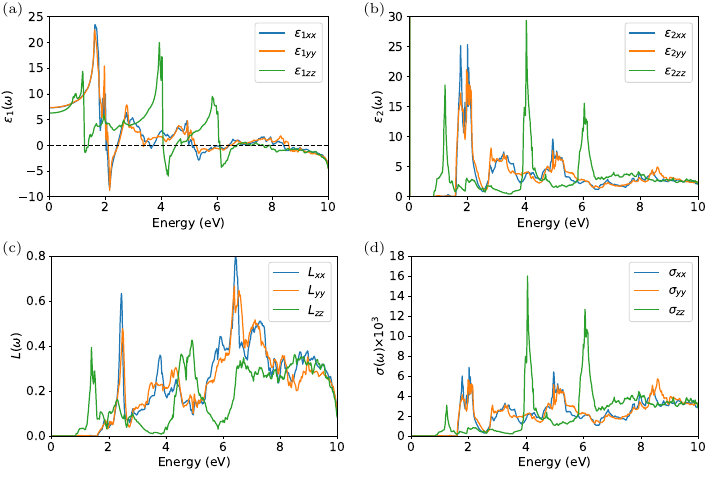} 
 		\vspace{-0.9cm}
 		\caption{Optical properties of Rb$_2$Ni$_3$S$_4$. (a) real part of the dielectric function $\epsilon_1$$(\omega)$. (b) Imaginary part of the dielectric function $\epsilon_2$$(\omega)$. (c) Energy loss spectrum L$(\omega)$. (d) Optical conductivity $\sigma$$(\omega)$ [in Ohm$^{-1}$cm$^{-1}$].} 
 		\label{optical}
 	\end{figure*}

 In order to understand the optical behavior of Rb$_2$Ni$_3$S$_4$, we perform the optical calculations. The linear response of the system to electromagnetic radiation is associated to the interaction of photons with electrons. This relation can be defined using the complex dielectric function, $\epsilon(\omega)$ as follows:
\begin{equation}
\epsilon(\omega)=\epsilon_1(\omega)+i\epsilon_2(\omega)
\end{equation}
where, $\epsilon_1$ and $\epsilon_2$ are the real and imaginary part of the dielectric function, respectively \cite{he2006first,ghosh2017electronic}.
In accordance with the selection criteria, the momentum matrix elements between the occupied and unoccupied state are used to calculate the imaginary component of the dielectric function $\epsilon_2$, given by

\hspace{-0.65cm}$\epsilon_2(\omega)=\frac{4\pi^2 e^2}{m^2 \omega^2 V} \sum_{g, g'}|<kg|p|kg'>|^2$
\begin{equation}
 \delta(E_{kg}-E_{kg'}-\hbar\omega)
\end{equation}
where, \textit{V} is the volume of the unit cell, and $p$ is the momentum operator \cite{ambrosch2006linear,launay2004evidence}. Here,  \textit{E$_{kg}$} is the eigenvalue (approximated to the GGA eigenvalues) associated with eigenfunction \textit{$|kg>$}. Likewise, the real part of dielectric function $\epsilon_1$ is given by the Kramers-Kronig relation  \cite{toll1956causality} as
\begin{equation}
\epsilon_1(\omega)=1+\frac{2}{\pi}\int\frac{\epsilon_2(\omega')\omega'}{(\omega')^2-\omega^2}d\omega'
\end{equation}
Additionally, the optical conductivity $\sigma$ and electron loss function $L$ are given by 
\begin{equation}
\sigma(\omega)=\frac{-i\omega(\epsilon(\omega)-1)}{4\pi}
\end{equation}

\begin{equation}
L(\omega)=\frac{\epsilon_2}{\epsilon_1^2(\omega)+\epsilon_2^2(\omega)}
\end{equation}

%\begin{equation}
 %\hat{H}=\sum_{{<\vec r,\vec {r'}>} \atop \alpha, \beta }T_{\alpha\beta} C^\dag_{\vec r, \alpha}C_{\vec {r'}, \beta} +h.c
%\end{equation}

We utilized these equations for the optical property calculations of Rb$_2$Ni$_3$S$_4$. As observed from the electronic band structure calculations, Rb$_2$Ni$_3$S$_4$ is found to be semiconductor with a band gap of $\sim$0.67 eV. Thus, when this material is exposed to an incident photons, transitions of electrons are expected. In such transitions, photon excites the electrons from an occupied states in the valence region to the unoccupied states in the conduction region (also called interband transition) \cite{stepanjuga2022metathesis}.  To understand these phenomena, we consider the polarization components along  x, y, and z directions as our material is orthorhombic in nature. We analyze the real part of the dielectric function $\epsilon_1(\omega)$ \cite{hoat2019systematic}, as depicted in Fig. 4(a) which gives rise to electronic polarizability of the material under an applied electric field. $\epsilon_1(\omega)$ along x, y and z direction are denoted by $\epsilon_{1xx}$,  $\epsilon_{1yy}$, and $\epsilon_{1zz}$, respectively. 
The real part of the dielectric function $\epsilon_1(\omega)$ at zero frequency, which is connected to the material's band gap (also known as static dielectric constants $\epsilon_1$(0)), are found to be $\sim$7.5 along both the x and y directions, while the value along the z direction is $\sim$6.2. This corresponds to the refractive index which can be determined using the established Penn Relation  n(0) = $\sqrt{\epsilon_1(0)}$ \cite{penn1962wave}. This give rise to the static refractive index value of 2.74 along x and y direction while 2.49 along z direction, respectively. The obtained value shows that the material has optical anisotropy. In the energy range 0 to 1.3 eV for the x, y direction, and 0 to 1 eV for the  z direction, we observe a gradual increase in $\epsilon_1(\omega)$, which indicates that the material interaction effectively with photons. Notably, at 1.63 eV, 1.60 eV, and 1.16 eV for $\epsilon_{1xx}$, $\epsilon_{1yy}$, and $\epsilon_{1zz}$, respectively, sharp peaks are observed. These peaks correspond to the inter-band transition, indicating the photon-induced electron transitions from occupied valence bands to the unoccupied conduction bands. This means that the material can absorb light particularly well at that specific energy.

When the photon energy range exceeds 1.16 eV, the real part of the dielectric function $\epsilon_1(\omega)$ exhibits a remarkable characteristic: it adopts a negative value. This phenomenon is indicative of incident light being reflected by the medium. Consequently, the material demonstrates metallic attributes in this specific energy range \cite{shore2019retraction,kumar2021structural}.  Furthermore, we also observe sharp peaks at approximately 2.01 eV, 1.96 eV, and 3.93 eV for x, y and z direction, respectively. $\epsilon_2(\omega)$, the imaginary part of the dielectric function, reveals details about optical absorption in crystal. The absorption begins at 0.84 eV for the imaginary part of the dielectric function, which acts as the threshold energy associated with the occurrence of an optical band gap (see Fig. 4 (b)).  Beyond the threshold point, the curve increases rapidly. Notably, sharp peaks are seen along the x, y, and z directions, respectively, at 2.02 eV, 1.97 eV, and 1.22 eV. These absorption peaks may clearly be associated with transitions from the Ni (3\textit{d}) valence band to the S(3\textit{p}) conduction band based on the predicted PDOS (see Fig. 2 (b)). Additionally, within the energy range of 2.83 eV to 6.05 eV, various significant peaks are observed. These peaks correspond to inter-band transitions from the valence band to the conduction band \cite{kumar2012first,belbase2023electronic}. However, beyond the energy of 6.05 eV, the peaks begin to diminish, eventually disappearing at higher energy ranges. 

Another important optical property calculated in this work is the electron energy loss function L($\omega$). Fig 4 (c) shows the variation of L($\omega$) with photon energy. The L($\omega$) provides the information about the energy loss of a fast-moving electron traversing in the material, and is reciprocal of the imaginary part of the complex dielectric function \cite{wang1981first,zhang2007electronic}. The visible area of the loss spectrum exhibits a slight energy loss, which rises as photon energy increases. The identified peaks in the L($\omega$) spectra provide information about the material's plasma frequency. Whether a material exhibits metallic properties or dielectric behavior is described by the plasma frequency. Materials exhibit metallic [$\epsilon_1(\omega)$ $\textless$ 0] behavior below the plasma resonance frequency and dielectric [$\epsilon_1(\omega)$$\textgreater$ 0] behavior above this frequency \cite{kumar2021structural,hu2007first,kumar2017optoelectronic,sun2006first}. The maxima in L$_{xx}$, L$_{yy}$, and L$_{zz}$ are clearly visible in the energy loss spectra at approximately 6.44 eV, 6.41 eV, and 4.91 eV, respectively. The resonant energy loss is seen at 6.44 eV along x direction with a maximum value of 0.8. After these peaks, the spectrum starts to decline.  
Fig. 4 (d) shows the optical conductivity which depends on photon energy. This indicates that the Rb$_2$Ni$_3$S$_4$ exhibits optical activity in the range from 1.25 eV to 8.60 eV. The $\sigma(\omega)$ edge starts 0.84 eV, which corresponds to the calculated optical band gap of the material, and numerous peaks are observed in the spectrum. The maximum optical conductivity, $\sigma_{xx}$, $\sigma_{yy}$ and $\sigma_{zz}$ was found to be 6.5$\times$10$^3$ Ohm$^{-1}$cm$^{-1}$, 16$\times$10$^3$ Ohm$^{-1}$cm$^{-1}$ and 5.9$\times$10$^3$ Ohm$^{-1}$cm$^{-1}$, respectively, at a photon energy of 2.02 eV to 4.05 eV. These findings are consistent with the imaginary part of the dielectric function (Fig. 4 (b)).

\section{Conclusions}
We have investigated the electronic structure and optical properties of Rb$_2$Ni$_3$S$_4$ by means of density functional theory approach. The studied material is found to be a non-magnetic semiconductor with an indirect band gap of $\sim$0.67 eV. The Ni atoms form a kagome lattice in a two-dimensional plane resulting in a flat band located below the Fermi energy. From the optical properties calculations, the dielectric function including its real and imaginary parts, the loss function, and the optical conductivity has been obtained. 
We found that the material is optically active in the visible and lower ultraviolet region. We thus predict that Rb$_2$Ni$_3$S$_4$ could be a possible candidate for optoelectronics  similar to some reported materials \cite{celep2004optical,adachi1993optical}.

 \section*{Acknowledgments}
 This work was supported by a grant from UNESCO-TWAS and the Swedish International Development Cooperation Agency (SIDA) (award no. 21-377 RG/PHYS/AS\textunderscore G). The views expressed herein do not necessarily represent those of UNESCO-TWAS, SIDA or its Board of Governors. 
This work is supported also by the Brain Pool program (No. RS-2023-00304344), “Regional Innovation Strategy (RIS)” (No. 2023RIS-009) and the National Research Foundation of Korea (NRF-2023K2A9A1A01098919). M.P.G. thanks the University Grants Commission, Nepal for the Collaborative Research Grants (award no. CRG-78/79 S\&T-03), and acknowledges IFW-Dresden for providing the large-scale compute nodes to Tribhuvan University for scientific computations.
 G.B.A. thanks Nepal Academy of Science and Technology (NAST) for the PhD fellowship. The author thanks Manuel Richter, IFW-Dresden for careful reading of the manuscripts and suggestions. M.P.G. and G.B.A. thanks Ulrike Nitzsche for the technical assistance.
 
 \bibliography{iopart-num}

\end{document}